# Dependability Evaluation of Stable Diffusion with Soft Errors on the Model Parameters

Zhen Gao, Lini Yuan, Pedro Reviriego, Shanshan Liu, Fabrizio Lombardi

*Abstract*—Stable Diffusion is a popular Transformer-based model for image generation from text; it applies an image information creator to the input text and the visual knowledge is added in a step-by-step fashion to create an image that corresponds to the input text. However, this diffusion process can be corrupted by errors from the underlying hardware, which are especially relevant for implementations at the nanoscales. In this paper, the dependability of Stable Diffusion is studied focusing on soft errors in the memory that stores the model parameters; specifically, errors are injected into some critical layers of the Transformer in different blocks of the image information creator, to evaluate their impact on model performance. The simulations results reveal several conclusions: 1) errors on the down blocks of the creator have a larger impact on the quality of the generated images than those on the up blocks, while the errors on middle block have negligible effect; 2) errors on the self-attention (SA) layers have larger impact on the results than those on the cross-attention (CA) layers; 3) for CA layers, errors on deeper levels result in a larger impact; 4) errors on blocks at the first levels tend to introduce noise in the image, and those on deep layers tend to introduce large colored blocks. These results provide an initial understanding of the impact of errors on Stable Diffusion.

## I. INTRODUCTION

Since proposed by Google in 2017, Transformers have been widely applied in the fields of Natural Language Processing (NLP) and computer vision (CV) [1]-[6], and become the basic network of most models for high-performance. Image generation is one of the most popular artificial intelligence capabilities, and has the ability to create striking visuals from text descriptions which promotes the expansion of industry, including games, films, and arts. Stable Diffusion is a Transformer based generative model for text to image generation [7][8], and its release is considered a milestone in this field because it is the first open source model which can generate high quality images with lower requirements for computation and memory.

The Stable Diffusion model is composed of three parts, a text encoder, an image information creator, and an image decoder [7]. The text encoder is used to condition the model on text prompts. The image information creator applies the diffusion technique to resemble the input text and all visual information learned from the training data in the latent representations. Finally, the image decoder generates the image from the latent representations. The image information creator is the key component that applies the diffusion technique to merge the semantics of the text in the latent image. This module applies an UNet architecture [9], and multiple ResNets and Transformers are involved in each down block or up block. Such a complex model requires a large volume of memory for the parameters.

Hardware implementations of deep neural networks (DNNs) are prone to suffer *soft errors* and the property of correct operation of DNNs in the presence of errors is often referred to as dependability. One of the most common soft errors is the Single-Event Upset (SEU) in memories, which is caused by the impact of cosmic rays and other radioactive particles on electronic devices. These errors are likely to occur in computing systems at nanoscales because transistors with aggressively scaled feature sizes become more susceptible to unexpected charge or noise (e.g., as caused by radiation particles) [10]-[12]. Therefore, to guarantee the correct execution of networks in critical scenarios that require reliable operation, the dependability of DNNs under soft errors must be carefully examined.

The impact of soft errors on the model performance have been extensively studied for Convolutional DNNs [13]-[19]; these works have shown that an error rate of $10^{-7}$ degrades the task performance dramatically for floating point weights, and the models with fixed point weights are more robust to SEUs. However, the dependability of Transformers against soft errors has not been well studied. Our previous work in [20] was the first to study the impact of soft errors on the weights of Transformers by taking Bidirectional Encoder Representations from Transformers (BERT) for sentence emotion classification as a case study; the results reveal that there is a Critical Bit (CB) on which errors significantly affect the performance of the model. The extension work in [21] provides a more detailed analysis of the dependability of BERT with soft errors on different components of the model. In addition, [22] considers the ReRAM based computing-in-memory system and proposes an efficient protection scheme for Transformers against errors on the loaded weights. All these works only consider the Transformer blocks in NLP applications. To the authors'

This work is supported by Natural Science Funds of China (62171313) and is partially supported by the ACHILLES (PID2019-104207RB-I00) and the ITACA (PDC2022-133888-I00) projects funded by the Spanish Agencia Estatal de Investigacion (AEI).

Zhen Gao and Lini Yuan, are with the School of Electrical and Information Engineering, Tianjin University, Tianjin, China. Email: {zgao, lnyuan}@tju.edu.cn.

Pedro Reviriego is with ETSI de Telecomunicación, Universidad Politécnica de Madrid, Madrid 28040, Email: pedro.reviriego@upm.es.

Shanshan Liu is with the School of Information and Communication Engineering, University of Electronic Science and Technology of China, Chengdu, Sichuan 611731, China. Email: ssliu@uestc.edu.cn.

Fabrizio Lombardi is with Department of Electrical and Computer Engineering, Northeastern University, Boston, MA 02215, USA. Email: lombardi@eoe.neu.edu.

best knowledge, the dependability of Transformer-based image diffusion models has not been studied.

This paper studies the dependability of Stable Diffusion with soft errors on the key model parameters. We inject SEUs on the critical bit of the weights and examine their impact when affecting different down/up/middle blocks and different attention layers and types in each block. The rest of this paper is organized as follows. Section II briefly reviews the Stable Diffusion model and its image information creator. The simulation setup for error injection and dependability evaluation method are introduced in Section III, and the simulation results are presented and discussed in Section IV. Finally, the paper is concluded in Section V.

## II. BASICS OF THE STABLE DIFFUSION MODEL

### A. Stable Diffusion

As shown in Fig. 1, Stable Diffusion is composed of a text encoder, an image information creator and an image decoder. The text encoder is the same as in the Contrastive Language-Image Pretraining (CLIP) [23], which transforms the input text with $M$ words to a $M*W$ text embedding matrix $E_T$, where $W$ is the embedding dimension. The image information creator is a UNet based diffuser with two inputs, the text embedding matrix and an $N_L*N_L$ latent image. The latent image is initially a noise image. During the diffusion process, a conditioned latent image is generated in each step based on the guidance of the text embeddings, and is used to estimate the noise that should be removed from the input latent image for the next step. Under the control of the Scheduler, a latent image with text semantics is generated after $L$ iterations. The image decoder is the decoder in the Variational Autoencoder (VAE) [24]. It generates an $N*N$ final image based on the latent image. Since the latent image is significantly smaller than the final image ($N_L < N$), Stable Diffusion can generate images of high quality at high efficiency.

For Stable Diffusion 2.0, the model configuration is set as $M = 77$, $W = 1024$, $N_L = 96$ and $N = 768$. With such a scale, the storage of the parameters requires 5.4 GB with half-precision format (float16), and the UNet component, which is the most complex component, accounts for over 2/3 of the memory.

### B. UNet based Image Information Creator

As shown in Fig. 2, the UNet part includes several down sample blocks (DBs), a middle block and several up sample blocks (UBs). The DBs compress the latent image to a lower resolution, and the UBs recover the latent image back to a higher resolution. To prevent the UNet from losing important information by down sampling, short-cut connections are added between DBs and UBs at the same level. From top to bottom, the first 3 DBs are Cross Attention Blocks (CABs), and the last one is a Down ResNet Block (DRB) without attention. Correspondingly, the bottom UB is an Up ResNet Block (URB) without attention, the other three are CABs. The middle block is also a CAB. As shown in Fig. 3, a CAB consists of multiple ResNets and Transformers and a convolution-based down or up sampler, and a Transformer which includes a self-attention (SA) layer and a cross-attention (CA) layer and a feed forward network (FFN). The SA is used to extract semantics from a latent image, and the CA is used to condition the latent image with the text embeddings.

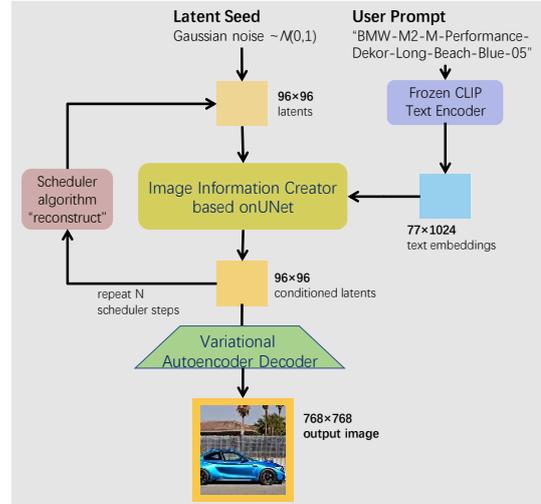

**Fig. 1.** Diagram of Stable Diffusion.

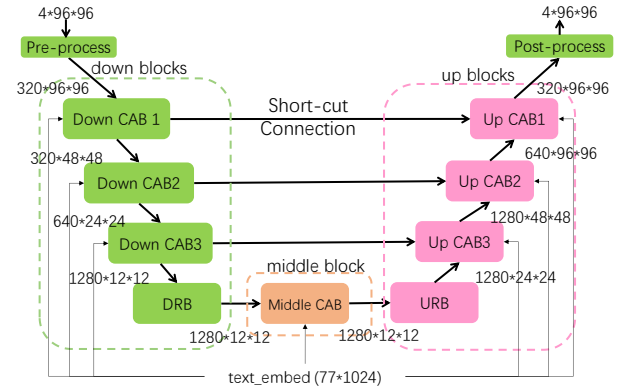

**Fig. 2.** Diagram of UNet based image information creator.

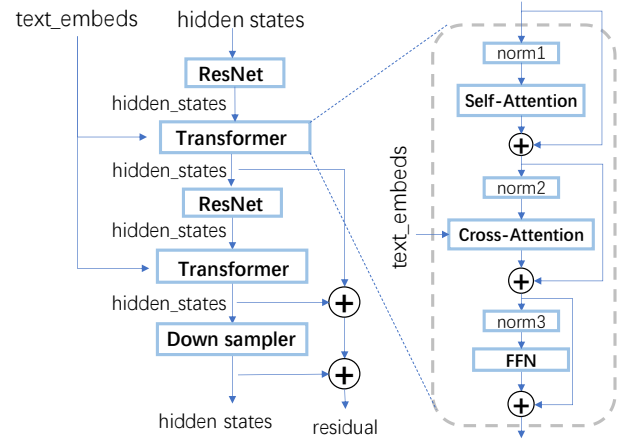

**Fig. 3.** Diagram of a CAB for DB.

In Stable Diffusion 2.0, each down side CAB includes two ResNets and two Transformers, each up side CAB includes three ResNets and three Transformers, and the CAB in the middle block includes two ResNets and one Transformer. In addition, each DRB is made of two ResNets, and each URB is made of three ResNets.

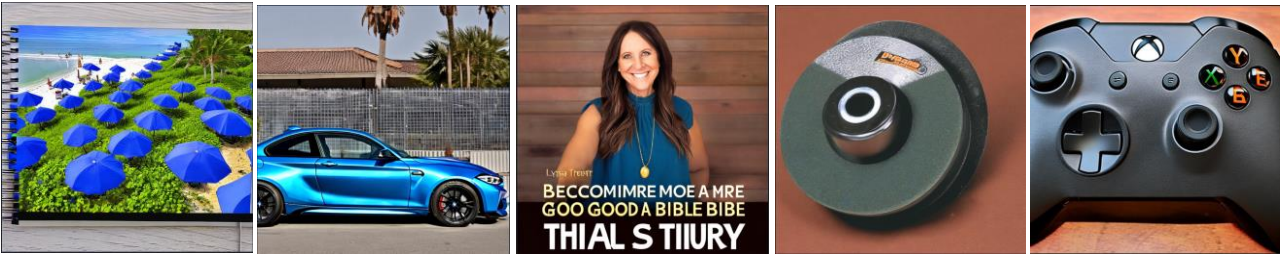

**Fig. 4.** Generated images for the first 5 prompts in the error-free case.

## III. SIMULATION SETUP AND EVALUATION METHOD

### A. Simulation Setup

We use the open-source code from [24] to implement Stable Diffusion 2.0 locally on a NVDIA 3070 GPU. The first five prompts from LAION2B-EN data set (set1) [26] are used as examples for the dependability evaluation of Stable Diffusion in this paper; they are described in Table I and the corresponding generated images in the error-free case are shown in Fig. 4. We apply *CLIP score* to evaluate the quality of the generated images [27]. First, the ViT image encoder in CLIP is used to transform the image into a 768*768 semantic embedding matrix $E_I$ [23]. Then the *cosine* similarity between the image embedding matrix and the text embedding matrix is calculated based on equation (1). In our implementation, the CLIP_score is calculated using the PyTorch function *torchmetrics.multimodal.clip_score*. With such metric, a higher score means the quality of the image is higher and the semantics of the image are closer to those of the prompt.

$$\text{CLIP\_score} = 100 \cdot max(0, \cos(E_I, E_T)) \quad (1)$$

It should be noted that the generated image may be different for the same prompt for different runs due to the random initial latent image. Therefore, to eliminate the influence of randomness on the dependability evaluation results, we use a fixed noise image as the input of the diffuser in all simulations. In this case, the CLIP scores for the five prompts are listed in Table I. The error-free CLIP score is around 30, and the average value is 30.22.

Table 1. First Five Prompts in LAION2B-EN (set1) and CLIP Scores

| Index | Prompt | CLIP score |
|---|---|---|
| 1 | Blue Beach Umbrellas, Point Of Rocks, Crescent Beach, Siesta Key - Spiral Notebook | 33.96 |
| 2 | BMW-M2-M-Performance-Dekor-Long-Beach-Blue-05 | 26.03 |
| 3 | Becoming More Than a Good Bible Study Girl: Living the Faith after Bible Class Is Over by Lysa TerKeurst Narrated by Lysa TerKeurst | 32.31 |
| 4 | "Dynabrade 52632 4-1/2" Dia. Right Angle Depressed Center Wheel Grinder | 28.25 |
| 5 | MANETTE XBOX ONE | 30.56 |

### B. Evaluation Method

As introduced above, we aim to evaluate the impact of SEUs on the model parameters in the quality of generated images. Since the UNet-based image information creator accounts for most of the complexity and the parameters of the model, the CABs are its key components and the Transformer is more complex than ResNets, then we focus on the evaluation of Transformers in CABs. Based on the principle of Transformer [1], in each SA or CA layer we have weight matrices $W_k$, $W_q$ and $W_v$ for the calculation of the Key, Query and Value matrices, $W_o$ for merging multiple heads, and $W_{f1}$ and $W_{f2}$ for the two fully-connection layers in the FFN. We analyzed the distribution of the float16 parameters in Transformers, and the average values of different bits are shown in Fig. 5. Since the magnitude of all the weights is smaller than 1, the 1st exponent bit is always 0. In addition, the 2nd exponent bit is 1 with probability of 68.7%, and all other bits can be 0 or 1 with approximately 50% probability.

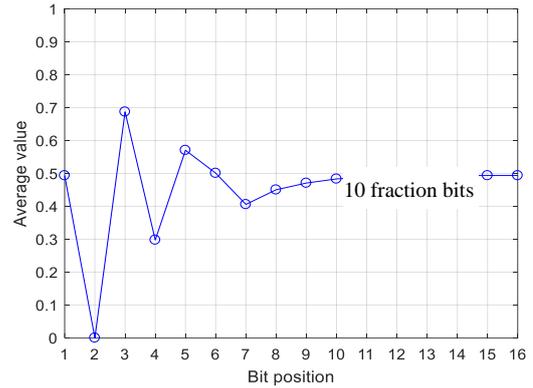

**Fig. 5.** Average value of different bits for parameters in the Transformers.

This is like the scenario in BERT as found in [21]. The four main conclusions of that paper are: 1) only SEUs on the 1st exponent bit may obviously degrade the Transformer performance because the 0→1 flip would introduce huge change of the intermediate results; 2) SEUs on other bits almost do not affect the task performance because 1→0 flips on the 2nd exponent are equivalent to pruning some weights and the random flips on other bits only introduce small changes to the intermediate results; 3) SEUs on the parameters for the Key and Query matrix in SA do not affect the task performance; 4) SEUs on $W_v$, $W_o$, $W_{f1}$ and $W_{f2}$ introduce obvious performance loss, and those on $W_v$ and $W_o$ have a similar effect. Therefore, in this paper we focus on $W_v$, $W_{f1}$ and $W_{f2}$ for the dependability evaluation, and on SEUs on the 1st exponent bit of a parameter. We conducted error injection on other bits to check that as in BERT, they have a negligible impact on the output images. Therefore, the first exponent bit is the critical bit, so for one SEU injection on a weight matrix in a SA or CA, we randomly select one item in the matrix and flip its 1st exponent bit. Then we run the model on the five prompts, and calculate the CLIP score. To eliminate the impact of weight positions to the result, the simulation is repeated 50 times, and the 250 CLIP scores are averaged to measure the effect of SEUs on that weight matrix. In this way, we can compare the effect of SEUs for different blocks, different Transformer layers and different attention types (CA or SA).

## IV. DEPENDABILITY EVALUATION RESULTS

In this section, the dependability of Stable Diffusion is evaluated under SEUs in the down blocks, up blocks and middle blocks, respectively. In each evaluation, we will first show the effect of SEUs with the change of CLIP scores, and then confirm the effect with examples of generated images. For simplicity, only the images for the second prompt (BMW car) are provided.

### A. Evaluation Results for Down Blocks

#### 1) Evaluation for SA and CA

The CLIP scores for SEUs on $W_v$ of the SA and CA layers for different DBs are shown in Fig. 6, where SA-T1 (or CA-T1) and SA-T2 (CA-T2) are for the SA (or CA) in the first and second Transformers in a DB. In general, the SEUs on $W_v$ decrease the CLIP score by 5 to 13 points. In addition, two observations can be established based on Fig. 6.

- **SA-Score**: The scores for SA in the two Transformers in a DB are similar, and do not change much for different DBs (the score for DB3 is slightly higher).
- **CA-Score**: The scores for CA decrease for the DB at a deeper level, and the scores for DB1 and DB2 are obviously higher than for SA.

Examples for the generated images for SEUs on SA and CA are provided in Figs. 7 and 8, respectively; since the images for T1 and T2 are similar, only one example for T1 is provided. The generated images with SEUs on different DBs show some obvious features.

- **SA-Feature**: Although the scores for SEUs on SA do not change much for different DBs, the generated images are obviously different. The image for DB1 is full of tiny noisy pixels; the image for DB2 includes blurry color strips; while for DB3, it is is composed of large blocks of colors.
- **CA-Feature**: The image for DB1 includes a complete car but the background details disappear; the image for DB2 includes a color block for the car and color strips for the background; for DB 3, it includes small color blocks for both the car and the background. This change is consistent with the decrease of the CLIP score.

Based on the above results, we can conclude that SEUs on CA have a smaller impact on the generated image than on SA. This is reasonable because SEUs on $W_v$ of CA only corrupt the Value matrix, but SEUs on $W_v$ of SA can corrupt the Key and Value matrices in CA. In addition, the reason for the SA-Feature and the CA-Feature can be explained based on the resolution of DBs at different levels. The corrupted pixels by SEUs on the shallower DBs corrupt all pixels of the latent image with lower resolution, which are then manifested as noise after the up sampling of UBs. SEUs on the deeper DB corrupt part of the pixels in the latent image at lower resolution, which are then manifested as larger color strips of blocks during the up-sampling process by the UBs.

#### 2) Evaluation for FFN

The CLIP scores for SEUs on the two fully-connection layers in the FFN ($W_{f1}$ and $W_{f2}$) for different DBs are shown in Fig. 9, in which FC1-T1 (or FC2-T1) and FC1-T2 (FC2-T2) are for the $W_{f1}$ (or $W_{f2}$) in the first and second Transformers in a down block. SEUs on the FFN decrease the CLIP score by about 9 to 12 points. In addition, the following observation can be made.

- **FNN-Score**: the scores for FC1 and FC2 in a Transformer are similar; the scores in the two Transformers in a DB are similar; and the scores do not change significantly for different DBs (the score for DB3 is slightly higher).

Fig. 10 provides the generated images with SEUs on the FFN. Since the images for FC1 and FC2 are similar, we only pick one example for FC1-T1 for three DBs. The main conclusion can be summarized as follows.

- **FFN-Feature**: Although the scores are similar, the features of the generated images are very different; the image for DB1 includes a colored car structure corrupted by noise; the image for DB2 includes blurry colored *brushstrokes*; the one for DB3 only includes large color blocks for both the car and the background. The reason for the trend is similar to SA and CA.

### B. Evaluation Results for Up Blocks

The CLIP scores for SEUs on the SA and CA layers for different UBs are shown in Fig. 11. As we can see, both the **SA-Score** and **CA-Score** hold for up blocks, and the only difference is that the CLIP scores are higher that the corresponding ones for DBs. Examples of generated images with SEUs on SA and CA are provided in Figs. 12 and 13, respectively. Both the **SA-Feature** and **CA-Feature** hold for UBs, but the only difference is that more information can be identified from the images due to the higher scores. For example, we can identify the outline of the car in the images for UB1 and UB2 in Fig. 13, which is significantly better than the corresponding ones in Fig. 7; compared with Fig. 8, more background details are included in the images for UB1 and UB2 in Fig. 14.

The CLIP scores for SEUs on FFN for different UBs are shown in Fig. 14, and the examples of generated images are provided in Fig. 15. Both the **FNN-Score** and **FFN-Feature** for DBs hold for UBs. The only small difference is that CLIP scores for UB2 are marginally higher than those for DB2. Therefore, the outline of the car can be identified in the image for UB2 in Fig. 15.

The smaller impact of SEUs on UBs can be explained by the short-cut connections between DBs and UBs. Corrupted latent images by SEUs on DBs tend to continue to affect the processing of UBs; for UBs, the impact of SEUs can be reduced by the non-corrupted results from DBs.

### C. Evaluation Results for Middle Blocks

The CLIP scores for SEUs on SA, CA and FFN of the middle block are listed in Table 2. As per the presented results, the scores are only slightly lower than the scores in the error-free case (30.22). We also checked the generated images, and found that they are almost the same as the error-free case.

Table 2. CLIP Scores for SEUs on Middle Block

|  | SA | CA | FC1 | FC2 |
|---|---|---|---|---|
| CLIP Score | 29.57 | 28.81 | 30.05 | 30.08 |

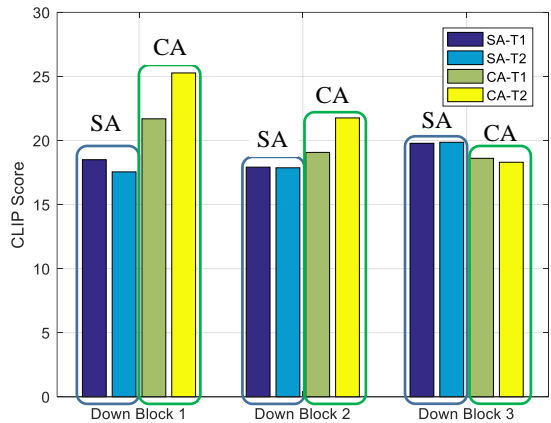

Fig. 6. CLIP Scores for SA and CA of Down Blocks

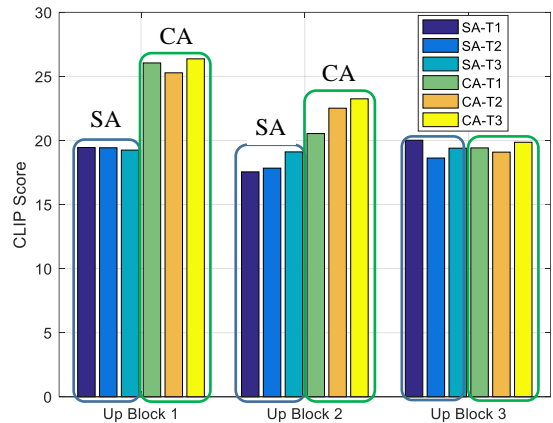

Fig. 11. CLIP Scores for SA and CA of Up Blocks

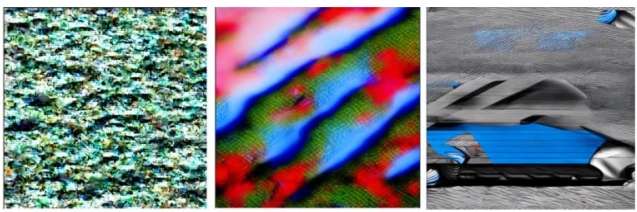

(a) DB1    (b) DB2    (c) DB3

Fig. 7. Generated Images with SEU on SA of Down Blocks

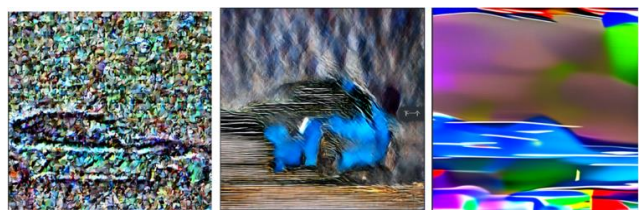

(a) UB1    (b) UB2    (c) UB3

Fig. 12. Generated Images with SEU on SA of Up Blocks

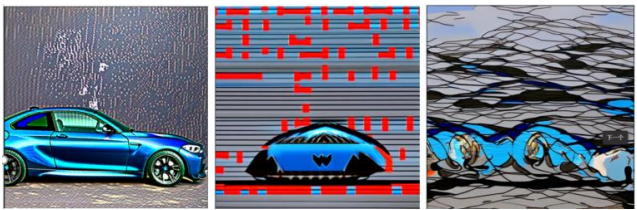

(a) DB1    (b) DB2    (c) DB3

Fig. 8. Generated Images with SEU on CA of Down Blocks

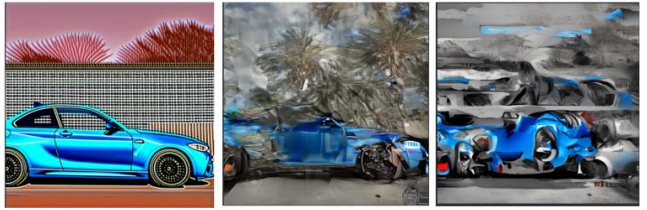

(a) UB1    (b) UB2    (c) UB3

Fig. 13. Generated Images with SEU on CA of Up Blocks

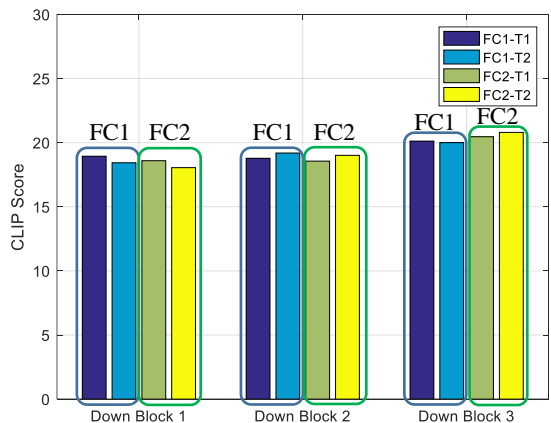

Fig. 9. CLIP Scores for FFN of Down Blocks

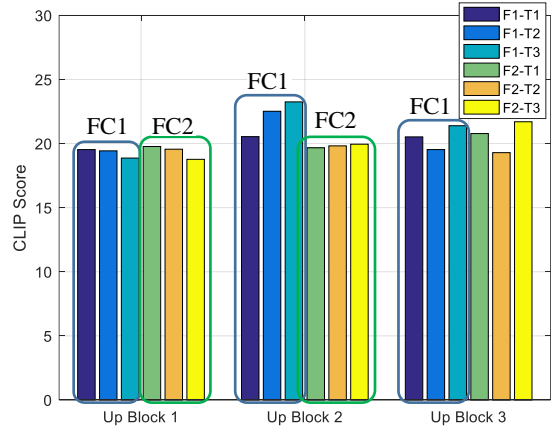

Fig. 14. CLIP Scores for FFN of Up Blocks

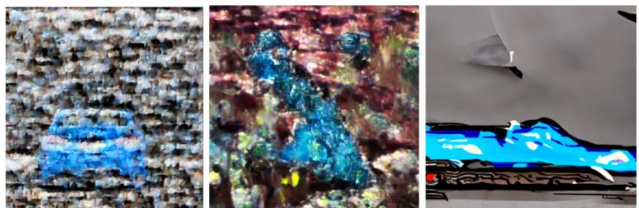

(a) DB1    (b) DB2    (c) DB3

Fig. 10. Generated Images with SEU on FFN of Down Blocks

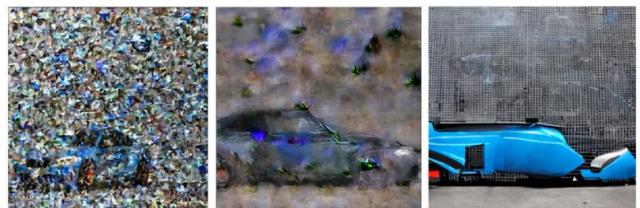

(a) UB1    (b) UB2    (c) UB3

Fig. 15. Generated Images with SEU on FFN of Up Blocks

## D. Effect of SEUs on Different Bits

In previous simulations, we assumed that SEUs only occur on the $1^{st}$ exponent bit. In this section, we validate such setting by injecting SEUs on other bits. In one SEU injection, we randomly select an item in $W_v$ in SA of the first Transformer in DB1, flip its $x$-th bit, run the model for the BMW prompt, and calculate the CLIP score for the generated image. This simulation is repeated 50 times, and the average score is considered as the effect of SEUs on bit $x$. The results are shown in Fig. 16; only SEUs on the $1^{st}$ exponent bit cause obvious decrease of the CLIP score, which confirms the assumptions for the above experiments.

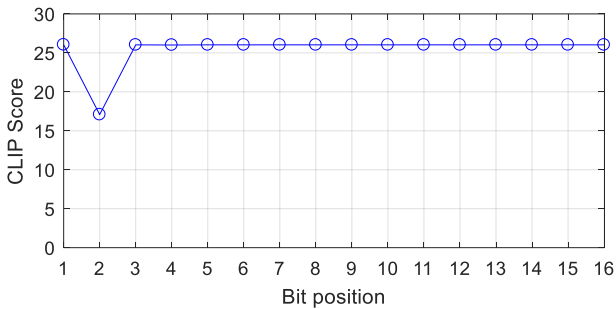

Fig. 16. Effect of SEUs on different bits (SA-T1 for BMW)

## V. CONCLUSION

Stable Diffusion is a successful generative model for image generation based on text prompts. Its key component is the UNet based image information creator, which conditions the image diffusion process with the text prompt. Each down sampler and up sampler block of UNet includes multiple ResNets and Transformers. In this paper, we have studied the dependability of Stable Diffusion in the presence of SEUs, which are one of the most frequent type of errors in nano-scaled memories. Specifically, we have injected SEUs on the parameters of several key parts of the Transformers, shown the impact of SEUs on the CLIP score of the generated images, and provided some examples for the corrupted images. Some interesting characteristics are drawn based on the simulation results.

- SEUs on self-attention (SA) layers cause a larger decrease of the image quality than those on cross-attention (CA) layers.
- SEUs on deeper blocks tend to introduce larger color blocks in the generated images.
- The impact of SEUs on up blocks is smaller than that for down blocks, and SEUs on the middle block only have a small impact.

A brief explanation is also provided throughout the paper by analyzing the features and architecture of the Stable Diffusion model.